\documentclass[12pt]{iopart}
\usepackage{iopams}

\usepackage{euscript,amssymb}

\newcommand{\be}[1]{\begin{equation}\label{#1}}
\newcommand{\ee}{\end{equation}}
\newcommand{\ba}[1]{\begin{eqnarray}\label{#1}}
\newcommand{\ea}{\end{eqnarray}}
\newcommand{\rf}[1]{(\ref{#1})}

\newcommand{\ov}{\overline}

\newcommand{\dd}{\dagger}

\def\p{\partial}
\def\a{\alpha}
\def\b{\beta}
\def\d{\delta}

\def\e{\varepsilon}

\def\vfi{\varphi}

\newcommand{\cD}{\mathcal{D}}

\newcommand{\cL}{\mathcal{L}}

\newcommand{\cR}{\mathcal{R}}

\newcommand{\ra}{\rangle}
\newcommand{\la}{\langle}

\def\wt{\widetilde}

\newcommand{\tb}[1]{\textbf{#1}}

\begin{document}

\title[SUSY transformations with complex factorization constants]%
{SUSY transformations with complex factorization constants. Application to spectral
singularities}

\author{Boris F. Samsonov}

\address{Physics Department, Tomsk State University, 36 Lenin Avenue,
634050 Tomsk, Russia}

\eads{\mailto{samsonov@phys.tsu.ru}}
%\centerline{\small\today}

\begin{abstract}
Supersymmetric (SUSY) transformation operators with complex
factorization constants are analyzed as operators acting in the Hilbert
space of functions square integrable on the positive semiaxis.
Obtained results are applied to Hamiltonians possessing spectral
singularities which are non-Hermitian SUSY partners of selfadjoint operators.
A new regularization procedure for the resolution of the identity
operator in terms
of continuous biorthonormal set of the non-Hermitian
Hamiltonian eigenfunctions is proposed.
It is also argued that, if the binorm of continuous spectrum
eigenfunctions is interpreted in the same way as the norm of similar
functions in the usual Hermitian case, then one can state that
 the function corresponding to a spectral singularity has zero binorm.
\end{abstract}
%\pacs{03.65.Xp, 03.65.Ca, 03.65.Ud, 03.67.Lx}
%\maketitle

\vspace{1em}
{\em Introduction.}\quad
In many cases the
spectrum of non-selfadjoint operators
(e.g. Hamiltonians) differ from the spectrum of
selfadjoint ones by two essential features.
These are (i) the possible
presence of exceptional points and (ii) the possible presence of spectral
singularities.
Note that exceptional points appear already in the finite dimensional case
(matrices) whereas spectral singularities
are a characteristic feature of Hamiltonians possessing a continuous
spectrum and, hence, they are impossible for finite dimensional
operators.
We think that just by this reason
 non-selfadjoint operators with spectral singularities are
studied in much less details.

Recently, one can notice a growing interest to Hamiltonians possessing
a real spectrum and
spectral singularities \cite{collection,Most,ACS,Gus}.
Probably it is due to a remark that they may produce a resonance-like
effect in some experiments \cite{Most}.
One can notice a contradiction in recently published results.
In particular, some authors claim that for a Hamiltonian possessing a
spectral singularity no any resolution of the identity operator is possible
\cite{collection,Most}.
From the other hand there is a thorough analysis of
 an exactly solvable complex potential
defined on the real axis where the authors prove that there always exists
a set of test functions for which the resolution of the identity operator over a
biorthonormal set of eigenfunctions takes place \cite{ACS}.

Note also that, as shown in \cite{my}, the method of supersymmetric
quantum mechanics (SUSY QM, for a review see e.g. \cite{Bogdan})
may be very useful for studying both exceptional points
and spectral singularities since non-Hermitian Hamiltonians possessing
these features may appear as SUSY partners of Hermitian Hamiltonians.
Another property of non-Hermitian Hamiltonians widely discussed in the
current literature is a possible real character of their spectrum
(for a review see \cite{Bender-review}).
Below we will consider only non-Hermitian Hamiltonians with a real
spectrum which are SUSY partners of Hermitian Hamiltonians.
This possibility is open after the publication of papers
\cite{Andr-real} where the authors
showed that SUSY
transformations may produce non-Hermitian Hamiltonians with real
spectra.

It happens that for a problem on a semiaxis, exceptional points appear if
the Jost solution with a real momentum
 is taken as the transformation function (see below).
 If we displace this
 real momentum from the real axis to the complex plane,
 the transformation function will correspond
 to a complex factorization constant but the spectral singularity disappears
 from the spectrum of the  non-Hermitian  Hamiltonian.
 In such a way one
 may realize a smooth path in the space of parameters, the non-Hermitian
 Hamiltonian $H$ depends on, leading from points where no any spectral
 singularity is present to a point where $H$ possesses a
 spectral singularity.
 This opens a way for regularizing the resolution of the identity
operator expressed in terms of continuous spectrum eigenfunctions of $H$
 in the
 point where the spectral singularity is present.
 To study this possibility in details, one
 has to consider the Hamiltonian in the vicinity of the spectral
 singularity in the space of parameters where no problems with the
 resolution of the identity operator is present and then consider a smooth limit
 to the singular point.

 To realize the above described procedure one needs to use
  SUSY transformations with {\it complex factorization constants}.
  Therefore we start with a thorough analysis of
 these transformations as operators
  acting in the Hilbert space of square integrable functions defined on the
  positive semiaxis.

  Below we will show that
a singularity inside the integral describing the resolution of the identity
operator
 is caused by a vanishing  bi-normalization coefficient
of corresponding
  continuous spectrum eigenfunction.
  Therefore, to regularize this resolution, it is sufficiently
  to shift the factorization constant
   by an infinitesimal value from the real axis to the complex
  plane
in the normalization factor of continuous spectrum eigenfunctions.
 Moreover, below we explain what we mean by the bi-normalization coefficient
 of the continuous spectrum eigenfunction and arrive at the conclusion that
 the function corresponding to the spectral singularity has zero binorm.

{\em SUSY transformations
of a real scattering potential with complex factorization constants.}\quad
Let we are given a Hermitian scattering Hamiltonian $h_0$
with a real valued potential $v_0(x)=v_0^*(x)$
\be{h0}
h_0=h_0^*=-\p_x^2+v_0(x)\,,\quad x\in\cR_+:=[0,\infty)\,.
\ee
It is initially
defined on the domain $\cD_{h_0}\subset\cL^2$
in the space $\cL^2:=\cL^2(\cR_+)$ of functions square
integrable on the non-negative semiaxis $\cR_+$.
As the domain $\cD_{h_0}$ one can choose a set of finite
twice continuously differentiable functions $\psi(x)$ with the
 Dirichlet boundary condition at the
origin, $\psi(0)=0$.
In this case the closure of $h_0$ is self-adjoint in $\cL^2$ (see, e.g., \cite{Kost_Sarg})
which we will denote by $\ov h_0=\ov h_0^\dag$ and its domain by
$\cD_{\ov h_0}$.

For simplicity we will assume also that $h_0$ has a purely continuous
spectrum,
\be{h0psi}
h_0\psi_k=E_k\psi_k\,,\ E_k=k^2\,,\ k\ge0\,.
\ee
Functions $\psi_k(x)$
(generalized eigenfunctions, see e.g. \cite{Gel-Shil})
 satisfy the Dirichlet boundary condition at the
origin, $\psi_k(0)=0$,
and asymptotic condition at $x=\infty$.
We would like to emphasize that they are assumed to be real valued,
\be{Repsi}
\psi_k^*(x)=\psi_k(x)\,,\quad k\ge0\,.
\ee
Moreover,
they form an orthonormal basis
(in the sense of distributions) in the space $\cL^2$
\be{ort-complet}
\la\psi_k|\psi_{k'}\ra:=\int_0^\infty\!\!\!dx\,\psi^*_k(x)\psi_{k'}(x)=
\d(k-k')\,,\quad
\int_0^\infty dk|\psi_k\ra\la\psi_k|={\mathbf 1}
\ee
and realize the spectral decomposition of $h_0$ which is just the colsure
$\ov h_0$ of the differential operator $h_0$ introduced above,
\be{h0-spectr}
\ov h_0=\int_0^\infty dk k^2|\psi_k\ra\la\psi_k|\,.
\ee
Note that $\psi_k(x)\notin\cL^2$ but they belong to a wider
space of functions defining regular functionals over a set of test
functions.
The action of the operator $h_0$ is naturally extended to the
space of these functionals.
 The interested reader may find more information about rigged
Hilbert spaces in \cite{Gel-Shil}.

Operator \rf{h0-spectr} is defined on a wider domain
$\cD_{\ov h_0}$ than $h_0$.
It consists of all functions $\psi\in\cL^2$ such that
\[
\|\ov h_0\psi\|^2:=\la\ov h_0\psi|\ov h_0\psi\ra
=\int_0^\infty\!\! dk\, k^4\la\psi|\psi_k\ra\la\psi_k|\psi\ra<\infty\,.
\]
Operator $\ov h_0$ presents the minimal closed and self-adjoint extension of $h_0$.

Operator \rf{h0} and a non-Hermitian Hamiltonian
\be{H}
H=-\p_x^2+V(x)
\ee
together with its adjoint $H^\dd=-\p_x^2+V^*(x)$ are related via
intertwining relations
\be{inter}
Lh_0=HL\,,\quad h_0L^\dd=L^\dd H^\dd\,.
\ee
Using the real valued character of $v_0$ we obtain a complex conjugate form
of equations \rf{inter}
\be{cinter}
L^*h_0=H^*L^*\,,\quad
h_0{L^*}^\dd={L^*}^\dd {H^*}^\dd={L^*}^\dd H\,.
\ee
Here we used the property $H^*=H^\dd$ which we assume to hold.
We would like to emphasize that $L^\dd$
 here should be understood as formally (Laplace) adjoint to $L$.
Below we will describe domains of the operators
$L$, $L^*$ and $L^\dd$ as differential operators acting in
the space $\cL^2$ and give their closed extensions.

In the simplest case, that we shall consider below, operator $L$ is a
first order differential operator
\be{L}
L=-\p_x+w
\ee
 with $w=w(x)=[\log u(x)]'$
being a complex valued
{\it superpotential}
defined with the help of a complex valued solution
$u=u(x)\ne0$ $\forall x\in(0,\infty)$ of the
differential equation $h_0u=\alpha u$ with $\alpha$ being, in general,
a complex factorization constant.
The constant $\alpha$ participates at the factorization of the Hamiltonians
$h_0$ and $H$
\be{factor}
{L^*}^\dd L=h_0-\alpha\,,\quad
L{L^*}^\dd=H-\alpha\,.
\ee
Complex conjugate form of these relations may also be useful
\be{factor-c}
{L}^\dd L^*=h_0-\alpha^*\,,\quad
L^*{L}^\dd=H^\dd-\alpha^*\,.\
\ee
Potential $V(x)$, defining Hamiltonian \rf{H},
 is expressed via the superpotential in the usual way,
$V(x)=v_0(x)-2w'(x)$.

Any solution $\vfi$ of the differential equation
\be{Hphi}
H\vfi_E=E\vfi_E\,,\quad E\ne\a
\ee
may be obtained by the action on a solution of the equation
$h_0\psi_E=E\psi_E$ with the operator $L$, $\vfi_E=L\psi_E$.
When $E=\a$ the function $\vfi_\a=1/u$ is a solution to equation \rf{Hphi},
$H\vfi_\a=\a\vfi_\a$. Another solution to this equation at $E=\a$,
$\wt\vfi_a$,
linearly independent with $\vfi_\a$, may be obtained from the equation
$\vfi_a(x)\wt\vfi{^{\,}}'_\a(x)-\vfi{^{\,}}'_a(x)\wt\vfi_\a(x)=1$.

In general, the
transformation $L$ violates boundary conditions.
The problem is simplified
essentially if $L$ transforms the eigenfunctions of the Hamiltonian $h_0$,
$\psi_k$,
to the
eigenfunctions of the Hamiltonian $H$, $\vfi_k$.
To guaranty this property we will
solve the boundary value problem defined by the equation \rf{Hphi} and
the boundary condition
\be{BC}
[\vfi{^{\,}}'_E(x)+\vfi_E(x) w(x)]_{x=0}=0\,.
\ee
The initial domain $\cD_H$ of the operator $H$
as an operator acting in the space $\cL^2$
consists of a set of finite twice
 continuously
differentiable functions $\vfi(x)$, satisfying condition
\rf{BC}.
The domain $\cD_{H^\dd}$ is defined as $\cD_{H^\dd}=(\cD_{H})^*$.
Note that $H^\dd$ defined on $\cD_{H^\dd}$ is not adjoint to $H$ in the
sense of the inner product.
Below we will give their closed extensions where this property will
take place.

Equations \rf{factor} and \rf{factor-c} are derived at the level of
differential expressions for all the operators involved in these
expressions.
Operators $h_0$, $H$ and $H^\dd$ have initial domains
in the space $\cL^2$ described above.
Using formulas \rf{factor} and \rf{factor-c}
 we can define $L$, $L^*$ and ${L^*}^\dd$
 as operators acting in the space $\cL^2$.
Evidently, operator $L$ \rf{L} is well defined on $\cD_{h_0}$.
Therefore we take $\cD_{h_0}$ as an initial domain for $L$ acting in
$\cL^2$.
Similarly, $\cD_{H}$ and  $\cD_{H^\dd}$ may be taken as initial domains
 for ${L^*}^\dd$ and ${L}^\dd$ respectively.

The functions $\wt\vfi_k=L\psi_k$, $k\ge0$ solve the equation $H\wt\vfi_k=k^2\wt\vfi_k$,
satisfy the condition \rf{BC}, and the asymptotic condition at infinity
and, hence, they are continuous spectrum eigenfunctions for $H$.
Similarly to $\psi_k$ they do not belong to the space $\cL^2$ but to a
wider space of functions defining regular functionals over a set of test
functions.
Similar to $h_0$
the operator $H$ is extended to the space of these functionals.
Note also that applying $L$ to $\psi_k\notin\cL^2$ we naturally extend the
action of the transformation operator to the space of functionals.
Below we will show that the functions $\vfi_k$ participate at the resolution  of
the identity operator.

The function $\vfi_\a=1/u$ which is a solution to Eq.
\rf{Hphi}, satisfies the condition \rf{BC}.
Therefore to have the spectrum of the operator $H$ real
for a complex $\a=-a^2$ with $\Im(\a)\ne0$,
we have to choose the function $u(x)$ such that
$1/u(x)\to\infty$ as $x\to\infty$.
This means that a good choice for $u(x)$ is the Jost solution
$f(k,x)$ for the Hamiltonian $h_0$,
$h_0f(k,x)=k^2f(k,x)$
which has the asymptotics
$f(k,x)\to\exp(ikx)$ as $x\to\infty$
at a fixed value of
 $k=-ia$, $a\in\mathbb{C}$.
 Thus,
\be{uas}
u(x)=f(-ia,x)\to\exp(a x)\,,\quad\a=-a^2\,,\quad x\to\infty
\ee
 with $a=d+ib$, $b\ne0$ and $d\le0$.
 For $d=0$, $u(x)$ oscillates at
infinity, i.e., it satisfies the asymptotic boundary condition
corresponding to a continuous spectrum eigenfunction.
As it will be shown
below, just in this case a spectral singularity appears in the spectrum of
$H$.
 For $d<0$, $u(x)\to0$ as $x\to\infty$ and the Hamiltonian $H$ has no
spectral singularities.
In both cases the operator $H$ has a purely
continuous spectrum,
$E=E_k=k^2$, $k>0$.
Below we will distinguish the case when $H$ has a spectral singularity
from that when its spectrum is regular.

{\em Regular case.}
Consider first the case $d<0$.
As it was already mentioned,
the functions $L\psi_k$, satisfy both the equation
\rf{Hphi} and the boundary condition \rf{BC}.
Therefore for $d<0$ the functions
\be{phik}
\vfi_k=N_k^{-1}L\psi_k\,,\quad N_k:=(k^2-\a)^{1/2}
\ee
form a continuous biorthonormal basis
(in the sense of distributions)
in $\cL^2$,
\be{biort}
\la\vfi^*_{k'}|\vfi_k\ra=
\int_0^\infty\!\!dx\,\vfi_k(x)\vfi_{k'}(x)=\d(k-k')\,,\quad
\int_0^\infty\!\!dk\,|\vfi_k\ra\la\vfi_k^*|=\mathbf{1}\,.
\ee
The first property follows from the first factorization relation in \rf{factor}
and the eigenvalue equation \rf{h0psi}.
The second property is a characteristic feature of a non-Hermitian
Hamiltonian with a purely continuous spectrum without spectral
singularities (see e.g. \cite{Naimark,Liantse}).
It may be justified by the following chain of equations
\[
\int_0^\infty\!\!dk\,|\vfi_k\ra\la\vfi_k^*|=
L\int_0^\infty\!\!dk\,\frac{|\psi_k\ra\la\psi_k^*|}{k^2-\a}(L^*)^\dag
=L(h_0-\a)^{-1}(L^*)^\dag
\]
\be{zzz3}
=(H-\a)^{-1}L(L^*)^\dag=(H-\a)^{-1}(H-\a)
=\mathbf{1}\,.
\ee
Here we used equations \rf{phik}, \rf{Repsi}, spectral decomposition
of the resolvent of $h_0$, first equation \rf{inter} and second equation
\rf{factor}.

Applying operator ${L^*}^\dd$ to \rf{phik},
using the factorization property \rf{factor} once again,
 and eigenvalue equation \rf{Hphi},
one obtains the transformation inverse to \rf{phik}
\be{vfik}
\psi_k=N_k^{-1}{L^*}^\dd\vfi_k\,.
\ee
%%%%%
Using the basis \rf{phik}, we obtain the spectral decomposition of $H$
which we will denote by $\ov H$
\be{Hspectr}
\ov H=\int_0^\infty\!dk\, k^2|\vfi_k\ra\la\vfi_k^*|\,.
\ee
From here one finds the operator adjoint to $\ov H$
\be{Hdds}
\ov H^\dd=\int_0^\infty\!dk\, k^2|\vfi_k^*\ra\la\vfi_k|\,.
\ee
Operator $\ov H$ is defined on a wider domain than $H$.
It consists of all functions $\vfi\in\cL^2$ such that
\[
\|\ov H\vfi\|^2=\la\ov H\vfi|\ov H\vfi\ra=
\int_0^\infty\!\!dk\,k^2\int_0^\infty\!\!dk'\,{k'}^2
\la\vfi|\vfi^*_{k'}\ra\la\vfi_{k'}|\vfi_k\ra\la\vfi^*_{k}|\vfi\ra<\infty\,.
\]

Evidently, operator $\ov H$ is densely defined by the relation \rf{Hspectr}.
Therefore $\ov H^\dd$ is closed.
Moreover, since  $\ov H=\ov H^{\dd\dd}$ this means that $\ov H$ is closed also
(see e.g. \cite{book-FA}).
Since $\ov H\vfi_k=H\vfi_k=k^2\vfi_k$ and
 $\ov H^\dd\vfi^*_k=H\vfi^*_k=k^2\vfi^*_k$,
 operators
 $\ov H$ and $\ov H^\dd$ are closed extensions of differential operators $H$ and $H^\dd$
respectively.

{\em{Closure of transformation operators in the space $\cL^2$ in the regular case.}}\quad
In the space $\cL^2$ we have two bases.
The functions $\psi_k$ form an
orthogonal basis and $\vfi_k$ form a biorthogonal one
(in the sense of distributions).
Transformation from
one basis to the other is realized by operators
\be{U}
U=\int_0^\infty\!\!dk\, |\vfi_k\ra\la\psi_k|\,,\quad U\psi_k=\vfi_k
\ee
and
\be{oU}
\wt U=\int_0^\infty\!\!dk\, |\psi_k\ra\la\vfi^*_k|\,,\quad \wt U\vfi_k=\psi_k\,.
\ee
They have the following property  $\wt UU=\mathbf1$, $U\wt U=\mathbf1$.
The first relation here is the usual resolution of the identity operator
over the orthonormal basis $\psi_k$ (see the second equation in
\rf{ort-complet})
whereas the second relation is nothing
but another form of the second equation in \rf{biort}.

Using these bases one can construct operators
\be{LB}
\ov L=\int_0^\infty\!\! dk\,N_k|\vfi_k\ra\la\psi_k|\,,\quad
\ov L^{\,\dd}=\int_0^\infty\!\! dk\,N^*_k|\psi_k\ra\la\vfi_k|
\ee
and
\be{LAB}
{{\ov L}{^{\,*}}}^\dd=\int_0^\infty\!\! dk\,N_k|\psi_k\ra\la\vfi^*_k|\,,\quad
\ov L^{\,*}=\int_0^\infty\!\! dk\,N_k^*|\vfi_k^*\ra\la\psi_k|\,.
\ee
Moreover, if we introduce the shifted versions of the Hamiltonians $h_0$ and $H$,
\be{g0}\fl
g_0:=h_0-\a=\int_0^\infty\!\! dk\, N_k^{2}|\psi_k\ra\la\psi_k|\,,\quad
g_1:=H-\a=\int_0^\infty\!\! dk\, N_k^{2}|\vfi_k\ra\la\vfi^*_k|
\ee
and their square roots
\be{SR-g0}
g_0^{1/2}=\int_0^\infty\!\! dk\, N_k|\psi_k\ra\la\psi_k|\,,\quad
g_1^{1/2}=\int_0^\infty\!\! dk\, N_k|\vfi_k\ra\la\vfi^*_k|\,.
\ee
then from \rf{U}, \rf{SR-g0}, \rf{LB}
and from \rf{oU}, \rf{g0}, \rf{LAB}
it follows that
\[
\ov L=Ug_0^{1/2}\,,\quad{{\ov L}^{\,*}}^\dd=\wt Ug_1^{1/2}\,.
\]
In case when $H$ is Hermitian, these relations reduce to polar decompositions
of the transformation operators obtained in \cite{my2000}.

From \rf{LB} we see that the domain of definition of $\ov L$
 consists of all functions $\psi\in\cL^2$ such
that
\[
\|\ov L\psi\|^2=\int_0^\infty\!\!dk\,\,N_k\int_0^\infty\!\!\,\,dk'N^*_{k'}
\la\psi|\psi_{k'}\ra\la\vfi_{k'}|\vfi_k\ra\la\psi_k|\psi\ra<\infty\,.
\]
Similarly,
from \rf{LAB} it follows that the domain of definition of ${{\ov L}^{\,*}}^\dd$
 consists of all functions $\psi\in\cL^2$ such
that
\[
\|{{\ov L}^{\,*}}^\dd\psi\|^2=\int_0^\infty\!\!dk\,|N_k|^2
|\la\psi|\vfi_k^*\ra|^2<\infty\,.
\]
Since $\ov L={\ov L}^{\dd\dd}$ these operators are closed
(see e.g. \cite{book-FA}).
Furthermore, it is easy to check that
\be{LLov}
\ov L\psi_k=L\psi_k=N_k\vfi_k\,,\quad
{{\ov L}^{\,*}}^\dd\vfi_k={{L}^{\,*}}^\dd\vfi_k=N_k\psi_k
\ee
meaning that they are closed extensions of the differential transformation
operators $L$ an ${L^*}^\dd$.

%%%%%%%%%%%%%%%%%%%%%%%%%%%%%%%%%%%%%%%%%%%%%%%%%%%%%%%%%%%%%%%%%%
{\em Spectral singularity.}\quad
The eigenfunctions $\psi_k$, $k\ge0$ of $h_0$ can be expressed via the Jost
solution $f(k,x)$ as (see e.g. \cite{Levitan})
\[\psi_k(x)=\sqrt{\frac{2}{\pi}}\frac{1}{|F(k)|}
\left[
F(-k)f(k,x)-F(k)f(-k,x)
\right],\quad k\ge0
\]
where $F(k)=f(k,0)$ is the Jost function for $h_0$.
Since $\ov h_0$ is self-adjoint, $F(k)$ does not vanish for any real $k\ne0$,
i.e.  $F(k)\ne0$ $\forall k\in \mathbb{R}\smallsetminus0$ \cite{Levitan}.
From here one finds
\be{LF}
L\psi_k(x)=\sqrt{\frac{2}{\pi}}\frac{1}{|F(k)|}
\left[
F(-k)Lf(k,x)-F(k)Lf(-k,x)
\right].
\ee
%%%%%%%%%%%%%%%%%%%%%%%%
%%%%%%%%%%%%%%%%%%%%%%%%
For any factorization constant $\a=-a^2$, $a=d+ib$ with $b\ne0$ and $d<0$, both
$Lf(k,x)\ne0$ and $Lf(-k,x)\ne0$ $\forall k\ge0$.
But for $d=0$ one has $\a=-a^2=b^2>0$ and since
$u(x)=f(-ia,x)=f(b,x)$, this implies that $Lf(b,x)=0$,
but $Lf(-b,x)\sim 1/u(x)$
and hence
$L\psi_{k=|b|}\sim 1/u(x)
%%%%%%%%%%%%%%%%%%%%%%%%%%%%%%%%%%%%%%%%%%%%%%%%%%%%%%%%%%%%%%%%%%%%%%%%%%%%%%%%%%%%%%%%
\to\exp(-ibx)$ as $x\to\infty$.
Therefore the function $\vfi_\a(x)= 1/u(x)$ has the asymptotics
$\vfi_\a(x)\to\exp(-ibx)$ as $x\to\infty$
(i.e., this is the Jost solution for $H$ at $k=-b$)
and at the same time,
as it was already mentioned, it satisfies the boundary condition
\rf{BC}.
By this reason the point $E=k^2=b^2$ belongs to the continuous spectrum of
$H$ and since the corresponding eigenfunction $\vfi_{k=|b|}(x)$
coincides (up to a nonzero constant)
 with the Jost solution for $H$,
 the point $E=k^2=b^2$,
 according to the definition (see e.g. \cite{Liantse,Gus,my}),
  is the spectral singularity for $H$.
By this reason
and by the definition of the Jost function as the value of the Jost
solution at the origin,
 the Jost function for $H$ vanishes at $k=-b$.
 Furthermore,
 the resolvent of $H$ grows in $\cL^2$ norm in the vicinity of $k=-b$
faster than in the absence of the spectral singularity \cite{Gus}
and a singularity appears
 in the resolution of the identity operator at $k=|b|$
  \cite{Liantse,ACS}.

%%%%%%%%%%%%%%%%%%%%%%%%%%%%%%%%%%%%%%%%%%

Another remarkable property of the function  $L\psi_{|b|}(x)$ is the value of
its binorm.
First we would like to stress that the notion of norm (or binorm) has a
sense if the normalization (or corresponding bi-normalization) integral
has a finite value which is not the case for the continuous spectrum
eigenfunctions.
Nevertheless, the usual physical terminology attaches a certain
significance to the term ``normalization of a continuous spectrum
eigenfunction''.
Namely, according to the first equation \rf{ort-complet}, one says that
the continuous spectrum eigenfunctions are ``normalized to the Dirac delta
function''.
Note that the Dirac delta is not a real function but a
distribution.
Thus, the above statement about the normalization of the
continuous spectrum eigenfunction $\psi_{k_0}$ should be understood in the sense of
distributions as a functional $\d(k-k_0)$ which, while acting on a test function
$\Phi(k)$,
gives the result $\Phi(k_0)$.
Just in a similar way,
taking into account the first formula \rf{biort},
one can say that
the continuous spectrum eigenfunction $\vfi_k$ for a Hamiltonian without
spectral singularities is ``bi-normalized to the Dirac delta function''.
Let us consider now what happens if the continuous spectrum eigenfunction
corresponds to a spectral singularity.

Taking into account factorization \rf{factor} and eigenvalue
equation \rf{h0psi}, one obtains
\be{0binorm}
\la(L\psi_k)^*|L\psi_{k'}\ra=\la\psi^*_k|{L^*}^\dd
L\psi_{k'}\ra=(k^2-\a)\d(k-k')
\ee
where the use of \rf{factor} has been made.
Evidently, this quantity vanishes at $k^2=\a=b^2$
and the binorm \rf{0binorm}, depending on the sign of $b$,
has a first order zero at either $k=b$ or $k=-b$
times the delta-function singularity which reduces
the right hand side of \rf{0binorm} to zero since the
delta-function belongs to a class of slowly increasing generalized functions.
We have to note that the right (and, hence, the left) hand side of
\rf{0binorm} should be understood in the sense of distributions.
This
means that acting by this functional on a test function of $k'$, one
obtains a function of $k$.
 If the test function is regular at ${k'}^2=\a=b^2$, the limit $k^2\to\a=b^2$ in
 \rf{0binorm} gives zero.
 In other words, the limit $k^2\to\a=b^2$ in \rf{0binorm}
 is zero functional defined on a set of test functions.
 In particular, as the set of test functions one can choose the space of
 continuous
square integrable functions $\Phi(k')$, $k'>0$.
 Actually, the set of test
functions is larger but here
  we leave open the question of how large it is.
We have to note that zero functional is regular.
This means that it may be presented
as an ordinary Lebesgue integral with a function ${\cal F}(k')$ which
is zero almost everywhere times a test function.
By this reason the value
of ${\cal F}(k'=|b|)$ does not affect the zero value of the
functional.
In particular, this function may even be undefined at $k'=|b|$
(see an example below).
Nevertheless, keeping the interpretation described above for the
normalization of the continuous spectrum eigenfunctions, we say that the
continuous spectrum
eigenfunction corresponding to the spectral singularity has zero binorm.
The final comment here is that this result is due to the vanishing
normalization factor $N_k^2=k^2-\a$ at $k^2=b^2$ if $\a=b^2$
present in \rf{0binorm}
but not to
the value of the bi-normalization integral \rf{biort} at $k=k'=\pm b$.

 Similar equation has been obtained for a non-Hermitian Hamiltonian
 defined in the Hilbert space of functions square integrable on the whole real line
  in \cite{ACS} (see Eq. (11) of that paper).
  This permits us to hope that the
the continuous spectrum eigenfunction
corresponding to a spectral singularity
   of a non-Hermitian Hamiltonian
has zero binorm
 not only for a class of
   Hamiltonians considered in this article but also for a more general
   case.

%%%%%%%%%%%%%%%%%%%%%%%%%%%%%%%%%%%%
{\em Resolution of the identity operator for a Hamiltonian with a spectral singularity.}\quad
For any $d<0$ the functions $\vfi_k$ realize the
resolution of the identity operator (see the second equation in
\rf{biort})
which we rewrite in the coordinate representation
\be{id-phi}
\int_0^\infty\!\!dk\,\vfi_k(x)\vfi_k(y)=
\int_0^\infty\!\!dk\,\frac{(L\psi_k)(x)(L\psi_k)(y)}{k^2-\a}=\d(x-y).
\ee
For $d=0$ the integrand contains a first order pole at $k^2=\a=b^2$. This
is just a characteristic feature of the spectral singularity.
In paper
\cite{Liantse} a regularization procedure is proposed for guarantying the
corresponding resolution of the identity operator.
In a recent paper \cite{ACS}, using a particular example,
 a thorough analysis of a similar regularization is given.
 Below, using the just established fact
 that the origin of this pole is the normalization
 factor for $\vfi_k$, we propose another regularization for this
 integral.

For $d=0$ ($\a=b^2$),
 instead of the functions $\vfi_k$ given in \rf{phik},
 let us consider the functions
 $\tilde\vfi_k=(k^2-b^2-i\e)^{-1/2}L\psi_k$
 with the sign of (infinitesimal)
 $\e$ possibly depending on the sign of $b$
 (see an example below)
 which differ
 from $\vfi_k$ by the normalization factor only and use them in the left
 hand (and the middle) part of equation \rf{id-phi}.
 The integrand in this equation is well defined $\forall\e\gtrless0$ now, but the
 right hand side is not the Dirac delta anymore.
 Using the property that
 this relation should be understood in the sense of distributions and the
 known fact (see e.g. \cite{Gel-Shil})
  that there always exists a set of test functions
 such that one can interchange the limit as $\e\to\pm0$ with the sign of
the integral, one restores the necessary behavior of the right hand side of
 \rf{id-phi} taking the limit $\e\to\pm0$ after the integral is calculated.
More precisely, we have first to apply the functional at the middle part of
\rf{id-phi} to a test function and after that take the
limit $\e\to\pm0$.
This statement may be additionally justified by the chain of
equations
\[\fl
\int_0^\infty\!\!dk\,|\tilde\vfi_k\ra\la{\tilde{\vfi}}_k^*|=
L\int_0^\infty\!\!dk\,\frac{|\psi_k\ra\la\psi_k^*|}{k^2-b^2-i\e}(L^*)^\dag
=L(h_0-b^2-i\e)^{-1}(L^*)^\dag
\]
\be{zxy}\fl
=(H-b^2-i\e)^{-1}L(L^*)^\dag=(H-b^2-i\e)^{-1}(H-b^2)
\to\mathbf{1}\,,\quad \e\to\pm0\,.
\ee
obtained in a similar way as the chain  \rf{zzz3}.
We note that two signs $\pm$ before $\e$ do not mean that both
signs are relevant. Only one sign leads to the desired result but it may
depend on the sign of $b$.

We have to note also that the above arguments have no strict mathematical
rigor and should be considered as an indication that this property may really take
place and probably may rigorously be proven following the lines similar to that which
one uses for proving the usual Parseval formula (see e.g. \cite{Levitan}).
%%%%%%%%                      %%%%%%%%%%%%%%%%%%%%%%%%%%%%%%

We would like also to mention that
the above described procedure is a typical method for
regularizing divergent integrals (see e.g. \cite{Gel-Shil}) but here we apply
it at the level of distributions.
Similar possibility for regularizing the resolution of the identity
operator
at the singular point is also mentioned in \cite{ACS} where the authors
note that the infinitesimal shifting the pole of an integrand from the
real axis to the complex plane may be considered as a procedure
alternative to a deformation of the integration contour.

%%%%%%%%               !!!!!!!!!!!!!!!

{\em Exactly solvable example.}\quad
Let us choose $v_0(x)=0$.
The Jost solution for $h_0$ is the simple
exponential $f(k,x)=\exp(ikx)$ so that $u(x)=\exp(ax)$, $a\in \mathbb{C}$,
$a\ne0$, and $\a=-a^2$.
The superpotential
is just a constant $w=a$ and the transformed Hamiltonian contains the
kinetic energy only, $H=-\p^2_x$, but the boundary condition at $x=0$ for
 equation \rf{Hphi} contains a complex number,
$\vfi{^{\,}}'(0)+a\vfi(0)=0$.
We recognize here an example of a non-Hermitian Hamiltonian with a
spectral singularity first proposed by Schwartz \cite{Schw} (see also
\cite{Gus}).
The functions $\vfi_k$ have the form
\be{ex-vfik}
\vfi_k=(k^2-\a)^{-1/2}\sqrt{\frac{2}{\pi}}
\left[
a \sin(kx)-k\cos(kx)
\right].
\ee
By the direct
calculation one finds the value of the integral
\[
\la(L\psi_k)^*|L\psi_{k'}\ra=(k^2+a^2)\d(k-k')
\]
which vanishes at $k^2=-a^2=b^2$ $\forall k'\ge0$ if $d=0$.
Thus, this is zero functional acting on a set of test functions of the variable
$k'$.
 The continuous spectrum eigenfunction
 for $k=|b|$ is $1/u=\exp(-ibx)$.
 If we tried to calculate the binorm of $1/u$
 using the first equation \rf{biort} for $k=k'=|b|$
 we will find that this quantity does not exist since the integral
of
 $\exp(-2ibx)$ over the interval $(0,\infty)$ does not converge.
This is just the illustration of the property discussed above that
 the function defining zero functional in the form of
 a usual Lebesgue integral is undefined for $k'=|b|$.
 Nevertheless, we say that the continuous spectrum eigenfunction
 corresponding to the spectral singularity has zero binorm
 in the sense described above.

%%%%%%%%%%%%%%%%            %%%%%%%%%%%%%%%%%                %%%%%%%%%%%%%%%%%%%%
Let us calculate now the integral in the left hand side of \rf{id-phi} for
$\a=b^2>0$.
 For this purpose we shift the parameter $b^2$
 in the denominator of the function \rf{ex-vfik} from the real axis
  to either the upper half or
 the lower half of the complex plane by an infinitesimal value $\e$,
$b^2\to b^2+i\e$.
We will see that the sign of $\e$ depends on the sign of $b$.
After some simple manipulations with trigonometric functions
and up to a term proportional to $\e^2$ in the denominator,
one obtains
for the integrand in \rf{zxy} the following approximation
\[
\frac\pi2\vfi_k(x)\vfi_k(y)\approx\cos(kx)\cos(ky)+
\left[
k^2+\left(\frac{\e}{2b}-ib\right)^2
\right]^{-1}
\]
\be{zz1}
\fl
\times
\left[
b^2\cos k(x+y)-ibk\sin k(x+y)+\frac{i\e}{2}(\cos k(x+y)+\cos k(x-y))
\right].
\ee
After integrating this expression with respect to $k$ over the interval $(0,\infty)$,
 the first term
gives just the desired Dirac delta function.
Thus, it remains to show that
all other terms give a vanishing result as
either $\e\to+0$ or $\e\to -0$.
Not that all integrals here are standard provided
$b\e>0$ and arguments of trigonometric functions are positive
(see formulas (3.723.2) and (3.723.3) in \cite{Gradshteyn}).
These integrals can easily be continued to the case $b\e<0$ and a negative
argument of cosine function,
\be{integrals}
\int_{0}^\infty \frac{\cos ax}{\beta^2+x^2}\, dx=
\pm\frac{\pi}{2\beta} e^{\mp |a|\beta}\,,\quad
\int_0^\infty \frac{x \sin cx}{\beta^2+x^2}\, dx=
\frac{\pi}{2}e^{\mp c\beta}\,,\quad c>0\,.
\ee
Here $\Im a=0$ and
the upper sign corresponds to $\Re\b>0$ and the lower sign to $\Re\b<0$.
Thus the desired integral may be approximated as
\be{zz2}\fl
\int_0^\infty\!\!\!dk\vfi_k(x)\vfi_k(y)\approx\d(x-y)+
\frac{e^{\mp z(x+y)}}{z}\left[
\pm b^2-ibz\pm \frac{i\e}{2}
\right]\!\!
\pm \frac{i\e}{2z}e^{\mp z|x-y|},\ z:=\frac{\e}{2b}-ib
\ee
with the upper sign for $b\e>0$ and the lower sign for $b\e<0$.
Replacing the parameter $z$
 in the middle term of the right hand side of
this equation by its value, one can see that this term
is equal to zero only for $b\e>0$.
For $b\e<0$ it survives in the limit $\e\to0$.
The last term is a regular functional acting on a test
function $\Phi(y)$, $y>0$, represented by the usual Lebesgue integral.
Splitting the integration interval in this integral with $b\e>0$
in two parts, $(0,x)$ and $(x,\infty)$,
one obtains
\[
\frac{i\e}{2z}\left[
e^{-zx}\int_0^x\!\!e^{zy}\Phi(y)dy+e^{zx}\int_x^\infty\!\! e^{-zy}\Phi(y)dy
\right],\quad \Re(z)>0\,.
\]
From here it follows that this is a decreasing function of $x$
 provided $\Phi(y)\to0$ as $y\to\infty$ and therefore for $\e\to0$ this
 term vanishes.
Thus, in the limit $\e\to0$
the right hand side of
formula \rf{zz2} reduces to the Dirac delta for a set of continuous test
functions vanishing at infinity and the sign of $\e$ coinciding with the
sign of $b$.

{\em Conclusion.}\quad
In this paper a careful analysis of SUSY transformations
with complex factorization constants
 as operator acting in the Hilbert space of square integrable functions
 defined on
 the positive semiaxis is given.
  Obtained results are applied to non-Hermitian Hamiltonians which are
  SUSY partners of self-adjoint operators.
Using the same arguments, which permit to say that the continuous spectrum
eigenfunctions of a Hermitian Hamiltonian are normalized to the Dirac delta
function, we arrived at the conclusion, that the continuous spectrum
eigenfunction of a non-Hermitian Hamiltonian
 corresponding to the spectral singularity has zero binorm.

  It is also shown that
a singularity inside the integral describing the resolution of the identity
operator
 is caused by the vanishing bi-normalization factor $(k^2-\a)$
of corresponding
  continuous spectrum eigenfunction with $\a=b^2>0$ at the singular point
  $k^2=b^2$.
   A new procedure for regularizing the resolution of the identity operator
   at the singular point is proposed.
   It consists of an infinitesimal shifting of the SUSY factorization constant
   from the real axis
   to the complex plane
   in the normalization factor of the corresponding scattering function.
   Obtained results are illustrated by the simplest example of a
   non-Hermitian Hamiltonian having the kinetic energy only but a complex
   boundary condition at the origin.
    Our results, thus, support the
   conclusion made in \cite{ACS} that the resolution of the identity operator for
   a non-Hermitian Hamiltonian possessing a spectral singularity may be
   constructed in contrast to the statement made in \cite{collection,Most}.

   Despite this positive result, we have to note that,
   since, as it has been noticed by Lyance \cite{Liantse}, there is
   no possibility for constructing a Hermitian operator equivalent to a
   non-Hermitian Hamiltonian with a spectral singularity, the relevance of
   such non-Hermitian Hamiltonians as fundamental objects in quantum
   mechanics is still doubtful.
   Nevertheless, for a non-Hermitian Hamiltonian nearly having a spectral
   singularity (and having no exceptional points) such a Hermitian
   operator exists and can be constructed.
   The scattering cross section for this Hamiltonian has the ordinary
   physical meaning and it may exhibit a resonance behavior near the singular
   point of its non-Hermitian counterpart \cite{My}.

   {\em Acknowledgment}.
   The author is grateful to A. Sokolov
   and to one of the referees
   for useful comments.
   The work is partially supported by
   Russian Science and Innovations
Federal Agency under contract 02.740.11.0238 and Russian Federal
Agency of Education under contract P1337.

\hspace{1em}

\end{document}